\documentclass[12pt]{article}
\usepackage{myart}

\normalbaselines
\input tcilatex
\begin{document}

\author{Yuri A. Rylov}
\title{Non-Riemannian geometry as a reason of the third modification of the
space-time geometry.}
\date{Institute for Problems in Mechanics, Russian Academy of Sciences,\\
101-1, Vernadskii Ave., Moscow, 119526, Russia.\\
e-mail: rylov@ipmnet.ru\\
Web site: {$http://rsfq1.physics.sunysb.edu/\symbol{126}rylov/yrylov.htm$}\\
or mirror Web site: {$http://gasdyn-ipm.ipmnet.ru/\symbol{126}%
rylov/yrylov.htm$}}
\maketitle

\begin{abstract}
The third modification of the space-time geometry is considered. (The first
modification is the spacial relativity, the second one is the general
relativity.) After the third modification of the space-time geometry the
motion of free particles become to be primordially stochastic
(multivariant). This circumstance forces one to construct a multivariant
dynamics. The multivariant dynamics is the classical dynamics in the
non-Riemannian space-time geometry. The multivariant dynamics explains
quantum effects without a reference to the quantum principles. Elimination
of quantum principles admits one to solve the main problem of relativistic
quantum theory: unification of the principles of relativity with the
nonrelativistic quantum principles.
\end{abstract}

\section{Introduction}

In the 20th century dynamics was developed mainly by means of modification
of the space-time model. Two essential modifications of the space-time
geometry were produced by Albert Einstein. The first modification solved the
problem of motion with large velocities. It is known as the special
relativity. The first modification led to Minkowski space-time as a result
of replacement of two invariants of the event space by one invariant.

The second modification is known as the general relativity. This
modification led to the Riemannian space-time geometry, which is a result of
influence of the matter distribution on the space-time curvature.

The third modification has not yet a short name. This modification led to
non-Riemannian space-time geometry (T-geometry). This modification of the
space-time geometry admitted one to describe quantum phenomena in the
framework of the classical dynamics (the quantum principles are not used).
In the non-Riemannian space-time geometry the motion of free particles is
multivariant (stochastic), and one needs a special multivariant dynamics,
which must be compatible with the space-time geometry. Note that the
space-time geometry in itself is single-variant (deterministic), whereas the
motion of particles in the single-variant space-time is multivariant
(stochastic). This is a corollary of the multivariance of the parallelism
concept.

In the Minkowski space-time geometry the deterministic motion is natural in
the sense, that there are no special reasons for the deterministic motion.
The natural stochastic motion in the Minkowski geometry is impossible. The
stochastic motion is artificial in the sense, that there must be some
special reason for the stochasticity.

In the non-Riemannian space-time the motion of any particle is stochastic
and there is no special reason for the stochasticity. In this sense the
stochastic motion is natural. We shall use a special term "multivariant
motion" for the natural stochastic motion. In some special case (the
particle of large mass) the multivariant motion degenerates into the
single-variant (deterministic) motion, which is a special case of the
multivariant motion.

The conception, generated by the third modification of the space-time
geometry, describes the quantum effects without a use of quantum principles.
It has the following properties.

\begin{enumerate}
\item It is a model conception, whereas the conventional quantum mechanics
is an axiomatic one.

\item It uses dynamical methods, which are not constrained by quantum
principles. The dynamical methods admit one to obtain results, which cannot
be obtained in the framework of conventional quantum mechanics: for
instance, incompatibility of the Copenhagen interpretation with the
formalism of quantum mechanics \cite{R2005,R2006}, existence of internal
degrees of freedom of the Dirac particle, which are described
nonrelativistically \cite{R2004,R2004a,R2005a,R2005b}, and the nature of the
pair production mechanism \cite{R2003}.

\item The fact, that the conception is \textit{more fundamental (primary)},
than the conventional quantum description, is of most importance.
\end{enumerate}

We can see this in the figure 1, where one presents some fundamental
conception, based on some primary (fundamental) propositions, which are
shown in the bottom. In the top we see experimental data, which may be
explained by means of corollaries of the primary propositions. It is
possible such a situation: the experimental data may be explained by means
of a set of corollaries, placed near the experimental data, without a direct
reference to primary propositions of the fundamental theory. In this case
the list of these propositions may be considered as primary propositions of
some theory (a curtailed theory). This curtailed theory may be considered to
be a self-sufficient theory, which does not need references to the
fundamental theory and does not use these references. The curtailed theory
contains more primary propositions, than the fundamental conception does,
because it contains corollaries of the primary propositions of the
fundamental theory, obtained in application to nonrelativistic phenomena.

If we do not know primary principles of the fundamental theory, we cannot
separate, what in the primary propositions of the curtailed theory is
conditioned by the primary propositions of the fundamental theory and what
is conditioned by the nonrelativistic character of the described phenomena.
In this case one may perceive the curtailed theory as a fundamental theory
with primary propositions other, than those of the fundamental theory. The
curtailed theory is axiomatic as a rule. Its application to explanation of
experimental data is easier and simpler, than the application of the
fundamental theory, because some corollaries of the fundamental theory are
contained in the curtailed theory in a ready form.

From the practical viewpoint the applications of the curtailed theory is
simpler, than the application of the fundamental theory. The curtailed
theory looks as a simpler theory, which is more convenient for explanation
of experimental data, than the fundamental theory. Besides, as a rule, the
curtailed theory is obtained earlier, than the fundamental theory, because
its primary propositions are nearer to experimental data and it is simpler,
than the fundamental theory. For instance, the axiomatic thermodynamics
(curtailed theory) had been constructed earlier, than the statistical
physics (kinetic theory), which is a fundamental theory with respect to
thermodynamics. As a rule, a transition from a known curtailed theory to the
corresponding unknown fundamental theory is difficult for perception of
researchers.

The nonrelativistic quantum theory is a curtailed theory, and there is a
fundamental theory for the nonrelativistic quantum theory. Unfortunately, at
the present stage of the science development, most researchers consider the
nonrelativistic quantum mechanics as a fundamental theory. The primary
propositions contain quantum principles, which are nonrelativistic. As a
result the problem of the relativistic quantum theory construction is
formulated as a join of nonrelativistic quantum principles with the
principles of relativity. Such a statement of problem is inconsistent.

The true statement of the problem is formulated as follows. One needs to
separate the nonrelativistic character of described phenomena from the
primary propositions of the theory. It means that one needs to construct a
fundamental theory, whose primary propositions are insensitive to the
character (relativistic or nonrelativistic) of the described phenomena.

\section{Why\ do\ \ we\ \ need\ \ T-geometry?}

In the thirties of the 20th century, one had discovered that free
microparticles move stochastically. Motion of free particles depends only on
the space-time geometry. To explain the stochasticity of the particle
motion, the space-time geometry was to possess the following properties.
Motion of a free particle in such a space-time geometry is primordially
stochastic (multivariant). Intensity of this stochasticity (multivariance)
is to depend on the particle mass. Such a geometry was not known till the
nineties of the 20th century.

In T-geometry the property of parallelism is intransitive, i.e. if we have
the relations $\mathbf{a}\parallel \mathbf{b}$ and $\mathbf{b}\parallel 
\mathbf{c}$ for vectors $\mathbf{a}$\textbf{,}$\mathbf{b}$\textbf{,}$\mathbf{%
c}$ , then, in general, vector $\mathbf{a}$ is not in parallel with vector $%
\mathbf{c}$. In reality, in T-geometry there are many vectors $\mathbf{b}%
_{1},\mathbf{b}_{2},...$, which are in parallel with vector $\mathbf{a}$,
but they are not in parallel between themselves. The intransitivity of
parallelism is connected with the multivariance of the parallelism, which
means, that there are many vectors (directions) $\mathbf{b}_{1},\mathbf{b}%
_{2},...$, which are in parallel with the vector (direction) $\mathbf{a}$,
but are not in parallel between themselves.

The world lines of free particles are shown in Figure 2. In the first case
the world line is shown in the Minkowski space-time. In the second case the
multivariant world line is shown in the space-time with intransitive
parallelism. In the first case we have the conventional single-variant
dynamics of the special relativity. In the second case one succeedes to
construct the corresponding multivariant dynamics (or a statistical
description). Such a dynamics is obtained, when the single particle (world
line) is replaced by a statistical ensemble of particles (world lines).

\section{Construction\ \ of\ \ T-geometry}

Any generalized geometry is obtained as a result of a deformation of the
proper Euclidean geometry. The proper Euclidean geometry is formalized, i.e.
any statement $\mathcal{S}$ and any geometrical object $\mathcal{O}$ is
represented in terms of the Euclidean world function $\sigma _{\mathrm{E}}$
in the form $\mathcal{S}\left( \sigma _{\mathrm{E}}\right) $ and $\mathcal{O}%
\left( \sigma _{\mathrm{E}}\right) $ respectively. (There is a theorem,
which states that it is always possible \cite{R2002}).\textit{\ The set of
all }$\mathbf{\ }\mathcal{S}\left( \sigma _{\mathrm{E}}\right) \ $\textit{and%
}$\mathbf{\ }\mathcal{O}\left( \sigma _{\mathrm{E}}\right) $ \textit{forms
the proper Euclidean geometry.} The world function \cite{S60} is defined by
the relation $\sigma \left( P,Q\right) =\frac{1}{2}\rho ^{2}\left(
P,Q\right) $, where $\rho \left( P,Q\right) $ is a distance between the
points $P$ and $Q$. It has the properties%
\begin{equation}
\sigma \left( P,P\right) =0,\qquad \sigma \left( P,Q\right) =\sigma \left(
Q,P\right) ,\qquad \forall P,Q\in \Omega  \label{a3.1}
\end{equation}

To obtain some generalized geometry $\mathcal{G}$, described by the world
function $\sigma $, it is sufficient to replace $\sigma _{\mathrm{E}}$ by $%
\sigma $ in all expressions $\mathcal{S}\left( \sigma _{\mathrm{E}}\right) $
and $\mathcal{O}\left( \sigma _{\mathrm{E}}\right) $. Then \textit{the set
of all}$\mathbf{\ }\mathcal{S}\left( \sigma \right) \ $\textit{and}$\mathbf{%
\ }\mathcal{O}\left( \sigma \right) \ $\textit{forms the} \textit{generalized%
} \textit{geometry}$\mathbf{\ }\mathcal{G}$, described by the world function 
$\sigma $.

The representation of the proper Euclidean geometry in the formalized ($%
\sigma $-immanent) form does not contain any theorems. All theorems are
replaced by definitions.

We shall explain this in the example of the cosine theorem, which states%
\begin{eqnarray}
\left\vert \mathbf{BC}\right\vert ^{2} &=&\left\vert \mathbf{AB}\right\vert
^{2}+\left\vert \mathbf{AC}\right\vert ^{2}-2\left( \mathbf{AB.AC}\right)
\label{a3.2} \\
&=&\left\vert \mathbf{AB}\right\vert ^{2}+\left\vert \mathbf{AC}\right\vert
^{2}-2\left\vert \mathbf{AB}\right\vert \left\vert \mathbf{AC}\right\vert
\cos \alpha  \nonumber
\end{eqnarray}%
where the points $A,B,C$ are vertices of a triangle, $\left\vert \mathbf{BC}%
\right\vert $, $\left\vert \mathbf{AB}\right\vert $, $\left\vert \mathbf{AC}%
\right\vert $ are lengths of the triangle sides and $\alpha $ is the angle $%
\angle BAC$. The relation (\ref{a3.2}) is the cosine theorem which is proved
on the basis of the axioms of the proper Euclidean geometry.

Using expression of the length of the triangle side $\mathbf{AB}$ via the
world function $\sigma $%
\begin{equation}
\left\vert \mathbf{AB}\right\vert =\sqrt{2\sigma \left( A,B\right) }
\label{a3.3}
\end{equation}%
we may rewrite the relation (\ref{a3.2}) for $\left\vert \mathbf{BC}%
\right\vert ^{2}$ in the form%
\begin{equation}
\left( \mathbf{AB.AC}\right) =\sigma \left( A,B\right) +\sigma \left(
A,C\right) -\sigma \left( B,C\right)  \label{a3.4}
\end{equation}%
This relation is a definition of the scalar product $\left( \mathbf{AB.AC}%
\right) $ of two vectors $\mathbf{AB}$\textbf{\ }and $\mathbf{AC}$, having
the common origin $A$. Thus, the theorem is replaced by the definition of a
new concept (the scalar product), which appears now not to be connected
directly with the concept of the linear space.

Another example the Pythagorean theorem for the rectangular triangle $ABC$
with the right angle $\angle BAC$. It is written in the form 
\[
\left\vert \mathbf{BC}\right\vert ^{2}=\left\vert \mathbf{AB}\right\vert
^{2}+\left\vert \mathbf{AC}\right\vert ^{2} 
\]%
In the formalized form (in T-geometry) we have a definition of the right
angle$\angle BAC$ instead of the Pythagorean theorem. In terms of the world
function this definition has the form. \textit{The angle }$\angle BAC$ 
\textit{is right, if the relation}%
\begin{equation}
\sigma \left( A,B\right) +\sigma \left( A,C\right) -\sigma \left( B,C\right)
=0  \label{a3.5}
\end{equation}%
\textit{takes place.}

Thus, the cosine theorem turns into the definition of the scalar product,
whereas the Pythagorean theorem turns into the definition of the right
angle. In a like way all theorems of the Euclidean geometry turn into
definitions.

Thus, we see that theorems of the proper Euclidean geometry are replaced by
definitions of T-geometry. The situation is very unusual and strange for
mathematicians, who cannot imagine any geometry without theorems, because
formulation and proof of theorems is the main work of geometers. Many of
geometers cannot accept the geometry without theorems, i.e. the formalized
form of the Euclidean geometry.

In reality, the Euclidean geometry is taken into account in the process of
the Euclidean geometry formalization. The conventional method of the
generalized geometry construction repeats all the Euclidean constructions at
other original axioms. The alternative method, based on the deformation
principle, does not need a repetition of all Euclidean constructions at the
obtaining of the generalized geometry. All Euclid's results are contained in
the formalized ($\sigma $-immanent) form of the Euclidean geometry.

\section{Parallelism of remote\ vectors. Multivariance of parallelism}

Scalar product $\left( \mathbf{P}_{0}\mathbf{P}_{1}.\mathbf{Q}_{0}\mathbf{Q}%
_{1}\right) $ of two remote vectors $\mathbf{P}_{0}\mathbf{P}_{1}$, $\mathbf{%
Q}_{0}\mathbf{Q}_{1}$ is defined by the $\sigma $-immanent relation 
\begin{equation}
\left( \mathbf{P}_{0}\mathbf{P}_{1}.\mathbf{Q}_{0}\mathbf{Q}_{1}\right)
=\sigma \left( P_{0},Q_{1}\right) +\sigma \left( P_{1},Q_{0}\right) -\sigma
\left( P_{0},Q_{0}\right) -\sigma \left( P_{1},Q_{1}\right)  \label{a4.1}
\end{equation}

Two vectors $\mathbf{P}_{0}\mathbf{P}_{1}$, $\mathbf{Q}_{0}\mathbf{Q}_{1}$
are linear dependent (collinear $\mathbf{P}_{0}\mathbf{P}_{1}||\mathbf{Q}_{0}%
\mathbf{Q}_{1}$), if the Gram determinant

\begin{equation}
\mathbf{P}_{0}\mathbf{P}_{1}||\mathbf{Q}_{0}\mathbf{Q}_{1}:\qquad \left\vert 
\begin{array}{cc}
\left( \mathbf{P}_{0}\mathbf{P}_{1}.\mathbf{P}_{0}\mathbf{P}_{1}\right) & 
\left( \mathbf{P}_{0}\mathbf{P}_{1}.\mathbf{Q}_{0}\mathbf{Q}_{1}\right) \\ 
\left( \mathbf{Q}_{0}\mathbf{Q}_{1}.\mathbf{P}_{0}\mathbf{P}_{1}\right) & 
\left( \mathbf{Q}_{0}\mathbf{Q}_{1}.\mathbf{Q}_{0}\mathbf{Q}_{1}\right)%
\end{array}%
\right\vert =0  \label{a4.2}
\end{equation}%
or%
\begin{equation}
\mathbf{P}_{0}\mathbf{P}_{1}||\mathbf{Q}_{0}\mathbf{Q}_{1}:\qquad \left( 
\mathbf{P}_{0}\mathbf{P}_{1}.\mathbf{Q}_{0}\mathbf{Q}_{1}\right)
^{2}=\left\vert \mathbf{P}_{0}\mathbf{P}_{1}\right\vert ^{2}\left\vert 
\mathbf{Q}_{0}\mathbf{Q}_{1}\right\vert ^{2}  \label{a4.3}
\end{equation}%
Here we see the definition of the linear dependence of two vectors, which
does not refer to the linear space. Again the theorem on necessary and
sufficient condition of linear dependence turns into definition of the
linear dependence.

Two vectors $\mathbf{P}_{0}\mathbf{P}_{1}$, $\mathbf{Q}_{0}\mathbf{Q}_{1}$
are in parallel $\mathbf{P}_{0}\mathbf{P}_{1}\uparrow \uparrow \mathbf{Q}_{0}%
\mathbf{Q}_{1}$, if 
\begin{equation}
\mathbf{P}_{0}\mathbf{P}_{1}\uparrow \uparrow \mathbf{Q}_{0}\mathbf{Q}%
_{1}:\qquad \left( \mathbf{P}_{0}\mathbf{P}_{1}.\mathbf{Q}_{0}\mathbf{Q}%
_{1}\right) =\left\vert \mathbf{P}_{0}\mathbf{P}_{1}\right\vert \cdot
\left\vert \mathbf{Q}_{0}\mathbf{Q}_{1}\right\vert  \label{a4.4}
\end{equation}

The set of such points $R$, that the vector $\mathbf{Q}_{0}\mathbf{R}$ is
collinear with the vector $\mathbf{P}_{0}\mathbf{P}_{1}$, forms the straight
(tube) $\mathcal{T}_{Q_{0};P_{0}P_{1}}$, passing through the point $Q_{0}$
collinear to the vector $\mathbf{P}_{0}\mathbf{P}_{1}$. 
\begin{equation}
\mathcal{T}_{Q_{0};P_{0}P_{1}}=\left( R|\mathbf{P}_{0}\mathbf{P}_{1}||%
\mathbf{Q}_{0}\mathbf{R}\right)  \label{a4.5}
\end{equation}%
If the straight passes in the four-dimensional space, the straight is, in
general, three-dimensional surface (multivariant straight). In the Minkowski
space-time the straight $\mathcal{T}_{Q_{0};P_{0}P_{1}}$ is one-dimensional
(single-variant), if the vector $\mathbf{P}_{0}\mathbf{P}_{1}$ is timelike ($%
\sigma \left( P_{0},P_{1}\right) >0$). The straight $\mathcal{T}%
_{Q_{0};P_{0}P_{1}}$ in the Minkowski space-time is three-dimensional
(multivariant), if the vector $\mathbf{P}_{0}\mathbf{P}_{1}$ is spacelike ($%
\sigma \left( P_{0},P_{1}\right) <0$). It is a reason, why one cannot
discover taxyons, when one searches them as one-dimensional spacelike lines.

If the space-time is deformed in such a way, that the world function $\sigma
_{\mathrm{d}}$ has the form%
\begin{equation}
\sigma _{\mathrm{d}}=\sigma _{\mathrm{M}}+d\left( \sigma _{\mathrm{M}%
}\right) ,\qquad d\left( \sigma _{\mathrm{M}}\right) =\left\{ 
\begin{array}{c}
\frac{\hbar }{2bc},\qquad \sigma _{\mathrm{M}}>\sigma _{0} \\ 
0,\qquad \sigma _{\mathrm{M}}<0%
\end{array}%
\right. ,  \label{a4.6}
\end{equation}%
the straight $\mathcal{T}_{Q_{0};P_{0}P_{1}}$ is always a three-dimensional
surface (multivariant straight). Here $\sigma _{\mathrm{M}}$ is the world
function of the Minkowski space-time, the quantities $\hbar $, $b$, $c$ are
constants.

Segment $\mathcal{T}_{\left[ P_{0}P_{1}\right] }$ of timelike straight $%
\mathcal{T}_{P_{0};P_{0}P_{1}}$ between the basic points $P_{0},P_{1}$ may
be presented in the form 
\begin{equation}
\mathcal{T}_{\left[ P_{0}P_{1}\right] }=\left( R|\sqrt{2\sigma \left(
P_{0},R\right) }+\sqrt{2\sigma \left( P_{1},R\right) }-\sqrt{2\sigma \left(
P_{0},P_{1}\right) }=0\right)   \label{a4.7}
\end{equation}%
A chain of such segments form the particle world line (tube)%
\begin{equation}
\mathcal{T}_{\mathrm{br}}=\dbigcup\limits_{i}\mathcal{T}_{\left[ P_{i}P_{i+1}%
\right] }  \label{a4.8}
\end{equation}%
The world line describe a free particle, if the vectors $\mathbf{P}_{i}%
\mathbf{P}_{i+1}$, $i=0,\pm 1,\pm 2,...$ are in parallel%
\begin{equation}
\mathbf{P}_{i}\mathbf{P}_{i+1}\uparrow \uparrow \mathbf{P}_{i+1}\mathbf{P}%
_{i+2}\qquad i=0,\pm 1,\pm 2,...  \label{a4.9}
\end{equation}%
and $\left\vert \mathbf{P}_{i}\mathbf{P}_{i+1}\right\vert =\mu $,$\ \qquad
ni=0,\pm 1,\pm 2,..$, $\mu $ is a geometrical mass of the particle. In the
case of Minkowski space-time and timelike vector $\mathbf{P}_{0}\mathbf{P}%
_{1}$, one obtains the one-dimensional (single-variant) straight line,
passing through the points $P_{0},P_{1}$.

In the case of the distorted space-time, described by the world function $%
\sigma _{\mathrm{d}}$, the world line of a free particle has the shape of a
multivariant broken tube. To describe such world tubes one needs a
multivariant dynamics.

\section{Multivariant dynamics}

Sir Isaac Newton had constructed his deterministic (single-variant) dynamics
for the Newtonian conception of space-time. The single-variant dynamics is
used for description of relativistic particles. However, for description of
multivariant world lines, one needs a multivariant dynamics. Multivariant
dynamics is used for description of particle motion in the Newtonian
space-time, or in the Minkowski space-time, when the initial conditions are
not known exactly. In this case one uses the concept of the statistical
ensemble.

We display in the example of free nonrelativistic particles, how the
statistical ensemble is introduced without a reference to the probability
theory, (i.e. only dynamically). The action $\mathcal{A}_{\mathcal{S}_{%
\mathrm{d}}}$ for the free deterministic particle $\mathcal{S}_{\mathrm{d}}$
has the form 
\begin{equation}
\mathcal{A}_{\mathcal{S}_{\mathrm{d}}}\left[ \mathbf{x}\right] =\int \frac{m%
}{2}\left( \frac{d\mathbf{x}}{dt}\right) ^{2}dt  \label{a5.1}
\end{equation}%
where $\mathbf{x}=\mathbf{x}\left( t\right) $.

For the pure statistical ensemble $\mathcal{A}_{\mathcal{E}\left[ \mathcal{S}%
_{\mathrm{d}}\right] }$ of free deterministic particles we obtain the action 
\begin{equation}
\mathcal{A}_{\mathcal{E}\left[ \mathcal{S}_{\mathrm{d}}\right] }\left[ 
\mathbf{x}\right] =\int \frac{m}{2}\left( \frac{d\mathbf{x}}{dt}\right)
^{2}dtd\mathbf{\xi }  \label{a5.2}
\end{equation}%
where $\mathbf{x}=\mathbf{x}\left( t,\mathbf{\xi }\right) $ is a 3-vector
function of independent variables $t,\mathbf{\xi =}\left\{ \xi _{1,}\xi
_{2},...\xi _{n}\right\} $. The variables (Lagrangian coordinates) $\mathbf{%
\xi }$ label particles $\mathcal{S}_{\mathrm{d}}$ of the statistical
ensemble $\mathcal{E}\left[ \mathcal{S}_{\mathrm{d}}\right] $. The
statistical ensemble $\mathcal{E}\left[ \mathcal{S}_{\mathrm{d}}\right] $ is
a dynamic system of hydrodynamic type. Note that the number $n$ of variables 
$\xi _{1,}\xi _{2},...\xi _{n}$ may be chosen arbitrary, but it is useful to
choose them in such a way, that the relations $\mathbf{x}=\mathbf{x}\left( t,%
\mathbf{\xi }\right) $ may be resolved in the form $\mathbf{\xi }=\mathbf{%
\xi }\left( t,\mathbf{x}\right) $. In this case we set $n=3$. The
statistical ensemble $\mathcal{E}\left[ \mathcal{S}_{\mathrm{d}}\right] $
realizes a multivariant description in the sense, that different values of $%
\mathbf{\xi }$ describe different world lines, determined by different
initial conditions.

If the particle $\mathcal{S=S}_{\mathrm{st}}$ is stochastic, dynamic
equations for the statistical ensemble $\mathcal{E}\left[ \mathcal{S}_{%
\mathrm{st}}\right] $ exist, whereas there are no dynamic equations for the
single stochastic particle $\mathcal{S}_{\mathrm{st}}$.

As a rule the statistical ensemble $\mathcal{E}\left[ \mathcal{S}\right] $
is considered as a derivative object. The basic object is a single dynamic
system. To construct the multivariant dynamics, we shall consider the
statistical ensemble $\mathcal{E}\left[ \mathcal{S}\right] $ as a basic
object, whereas the single particle is considered as a partial case of the
statistical ensemble with $\delta $-like initial data.

In the case of deterministic dynamic system $\mathcal{S}_{\mathrm{d}}$ the
dynamic equations, generated by the action for the single particle $\mathcal{%
S}_{\mathrm{d}}$, and those, generated by the statistical ensemble $\mathcal{%
E}\left[ \mathcal{S}_{\mathrm{d}}\right] $, are similar

\begin{equation}
\frac{d^{2}\mathbf{x}}{dt^{2}}=0  \label{a5.3}
\end{equation}%
The objects $\mathcal{S}_{\mathrm{d}}$ and $\mathcal{E}\left[ \mathcal{S}_{%
\mathrm{d}}\right] $ distinguish in the relation, that $\mathcal{S}_{\mathrm{%
d}}$ is described by one vector function $\mathbf{x}=\mathbf{x}\left(
t\right) $, whereas $\mathcal{E}\left[ \mathcal{S}_{\mathrm{d}}\right] $ is
described by many different functions $\mathbf{x}=\mathbf{x}\left( t,\mathbf{%
\xi }\right) $. In this case it is of no importance, which of two objects: $%
\mathcal{S}$ or $\mathcal{E}\left[ \mathcal{S}\right] $ is basic.

But for the stochastic particles the choice of basic object is important. If
the statistical ensemble $\mathcal{E}\left[ \mathcal{S}\right] $ is a basic
object, the dynamic equations exist always for it. The fact, that there are
no dynamic equations for a single stochastic particle is of no importance,
because the single particle is not a basic object, and the dynamics is a
dynamics of basic objects (statistical ensembles).

Thus, \textit{choosing}\ \textit{the}\ \textit{statistical}\ \textit{ensemble%
}\ \textit{as}\ \textit{a\ basic}\ \textit{object\ of\ dynamics},\textit{\
we\ may\ construct\ a\ multivariant\ dynamics}.

The statistical ensemble $\mathcal{E}\left[ \mathcal{S}_{\mathrm{st}}\right] 
$ of free \textit{stochastic} particles $\mathcal{S}_{\mathrm{st}}$ is a
dynamic system, described by the action 
\begin{equation}
\mathcal{A}_{\mathcal{E}\left[ \mathcal{S}_{\mathrm{st}}\right] }\left[ 
\mathbf{x,u}_{\mathrm{df}}\right] =\int \left\{ \frac{m}{2}\left( \frac{d%
\mathbf{x}}{dt}\right) ^{2}+\frac{m}{2}\mathbf{u}_{\mathrm{df}}^{2}-\frac{%
\hbar }{2}\mathbf{\nabla u}_{\mathrm{df}}\right\} dtd\mathbf{\xi }
\label{a5.4}
\end{equation}%
where $\mathbf{u}_{\mathrm{df}}=\mathbf{u}_{\mathrm{df}}\left( t,\mathbf{x}%
\right) $ is a diffusion velocity, describing the mean value of the
stochastic component of velocity, whereas $\frac{d\mathbf{x}}{dt}\left( t,%
\mathbf{\xi }\right) $ describes the regular component of the particle
velocity, and $\mathbf{x}=\mathbf{x}\left( t,\mathbf{\xi }\right) $ is the
3-vector function of independent variables $t,\mathbf{\xi =}\left\{ \xi
_{1,}\xi _{2},\xi _{3}\right\} $. The variables $\mathbf{\xi }$ label
particles $\mathcal{S}_{\mathrm{st}}$, substituting the statistical
ensemble. The operator 
\begin{equation}
\mathbf{\nabla =}\left\{ \frac{\partial }{\partial x^{1}},\frac{\partial }{%
\partial x^{2}},\frac{\partial }{\partial x^{2}}\right\}  \label{a5.5}
\end{equation}%
is defined in the coordinate space of $\mathbf{x}$. Note that the transition
from the statistical ensemble $\mathcal{E}\left[ \mathcal{S}_{\mathrm{d}}%
\right] $ to the statistical ensemble $\mathcal{E}\left[ \mathcal{S}_{%
\mathrm{st}}\right] $ is purely dynamic. The concept of probability is not
used. The character of stochasticity is determined by the form of two last
terms in the action (\ref{a5.4}) for $\mathcal{E}\left[ \mathcal{S}_{\mathrm{%
st}}\right] $ . For instance, if we replace $\mathbf{\nabla v}_{\mathrm{df}}$
by some function $f\left( \mathbf{\nabla v}_{\mathrm{df}}\right) $, we
obtain another type of stochasticity, which does not coincide with the
quantum stochasticity.

The action for the single stochastic particle is obtained from the action (%
\ref{a5.4}) for $\mathcal{E}\left[ \mathcal{S}_{\mathrm{st}}\right] $ by
omitting integration over $\mathbf{\xi }$. We obtain the action

\begin{equation}
\mathcal{A}_{\mathcal{S}_{\mathrm{st}}}\left[ \mathbf{x,u}_{\mathrm{df}}%
\right] =\int \left\{ \frac{m}{2}\left( \frac{d\mathbf{x}}{dt}\right) ^{2}+%
\frac{m}{2}\mathbf{u}_{\mathrm{df}}^{2}-\frac{\hbar }{2}\mathbf{\nabla u}_{%
\mathrm{df}}\right\} dt  \label{a5.6}
\end{equation}%
where $\mathbf{x}=\mathbf{x}\left( t\right) ,\ \ \mathbf{u}_{\mathrm{df}}=%
\mathbf{u}_{\mathrm{df}}\left( t,\mathbf{x}\right) $. However, this action
has only a symbolic sense, as far as the operator $\mathbf{\nabla }$ is
defined in some vicinity of the point $\mathbf{x}$, but not at the point $%
\mathbf{x}$ itself. It means, that this action does not determine dynamic
equations for the single particle $\mathcal{S}_{\mathrm{st}}$, and the 
\textit{particle appears to be stochastic}, although dynamic equations exist
for the statistical ensemble of such particles. They are determined by the
action (\ref{a5.4}) for $\mathcal{E}\left[ \mathcal{S}_{\mathrm{st}}\right] $%
. Thus, the particles described by the action for $\mathcal{E}\left[ 
\mathcal{S}_{\mathrm{st}}\right] $ are stochastic, because there are no
dynamic equations for a single particle. In the case, when the quantum
constant $\hbar =0$, the actions (\ref{a5.4}) for $\mathcal{S}_{\mathrm{st}}$
and (\ref{a5.2}) for $\mathcal{S}_{\mathrm{d}}$ coincide, because in this
case it follows from dynamic equation, that $\mathbf{u}_{\mathrm{df}}=0$.

Variation of action for $\mathcal{E}\left[ \mathcal{S}_{\mathrm{st}}\right] $
with respect to variable $\mathbf{u}_{\mathrm{df}}$ leads to the equation 
\begin{equation}
\mathbf{u}_{\mathrm{df}}=-\frac{\hbar }{2m}\mathbf{\nabla }\ln \rho ,
\label{a5.7}
\end{equation}%
where the particle density $\rho $ is defined by the relation 
\begin{equation}
\rho =\left[ \frac{\partial \left( x^{1},x^{2},x^{3}\right) }{\partial
\left( \xi _{1},\xi _{2},\xi _{3}\right) }\right] ^{-1}=\frac{\partial
\left( \xi _{1},\xi _{2},\xi _{3}\right) }{\partial \left(
x^{1},x^{2},x^{3}\right) }  \label{a5.8}
\end{equation}

Eliminating $\mathbf{u}_{\mathrm{df}}$ from the dynamic equation for $%
\mathbf{x}$, we obtain dynamic equations of the hydrodynamic type. 
\begin{equation}
m\frac{d^{2}\mathbf{x}}{dt^{2}}=-\mathbf{\nabla }U\left( \rho ,\mathbf{%
\nabla }\rho \right)  \label{a5.9}
\end{equation}%
\begin{equation}
U\left( \rho ,\mathbf{\nabla }\rho \right) =\frac{\hbar ^{2}}{8m}\left( 
\frac{\left( \mathbf{\nabla }\rho \right) ^{2}}{\rho ^{2}}-2\frac{\mathbf{%
\nabla }^{2}\rho }{\rho }\right)  \label{a5.10}
\end{equation}%
By means of the proper change of variables these equations can be reduced to
the Schr\"{o}dinger equation \cite{R99}.

However, there is a serious mathematical problem here. The fact is that the
hydrodynamic equations are to be integrated, in order they can be described
in terms of the wave function. The fact, that the Schr\"{o}dinger equation
can be written in the hydrodynamic form, is well known \cite{M26}. However,
the inverse transition from the hydrodynamic equations to the description in
terms of wave function was not known until the end of the 20th century \cite%
{R99}, and the necessity of integration of hydrodynamic equations was a
reason of this fact.

Derivation of the Schr\"{o}dinger equation as a partial case of dynamic
equations, describing the statistical ensemble of random particles, shows
that the$\ $\textit{wave}\textbf{\ }\textit{function}\textbf{\ }\textit{is}%
\textbf{\ }\textit{simply}\textbf{\ }\textit{a\ method}\textbf{\ }\textit{of}%
\textbf{\ }\textit{description}\textbf{\ }\textit{of}\textbf{\ }\textit{%
hydrodynamic}\textbf{\ }\textit{equations}\textbf{,\ }\textit{but}\textbf{\ }%
\textit{not}\textbf{\ }\textit{a}\textbf{\ }\textit{specific}\textbf{\ }%
\textit{quantum}\textbf{\ }\textit{object}, whose properties are determined
by the quantum principles. At such an interpretation of the wave function
the \textit{quantum\ principles\ appear\ to\ be\ superfluous}, because they
are necessary only for explanation, what is the wave function and how it is
connected with the particle properties. All remaining information is
contained in the dynamic equations. It appears that the \textit{quantum\
particle\ is\ a\ kind\ of\ stochastic\ particle}, and all its exhibitions
can be interpreted easily in terms of multivariant dynamics (in terms of the
statistical ensemble of stochastic particles).

The idea of that, the quantum particle is simply a stochastic particle, is
quite natural. It was known many years ago. However, the mathematical form
of this idea could not be realized for a long time because of the two
problems considered above (incorrect conception on the statistical ensemble
of relativistic particles and necessity of integration of the hydrodynamic
equations).

One can show, that quantum systems are a special sort of dynamic systems,
which could be obtained from the statistical ensemble of classical dynamic
systems by means of a change of parameters $P$ of the dynamic system by its
effective value $P_{\mathrm{eff}}$. In particular, the free uncharged
particle is described by an unique parameter: its mass $m$.

Statistical ensemble of free classical relativistic particles is described
by the action 
\begin{equation}
\mathcal{A}_{\mathcal{E}\left[ \mathcal{S}_{\mathrm{d}}\right] }\left[ x%
\right] =-\int mc\sqrt{g_{ik}\dot{x}^{i}\dot{x}^{k}}d\tau d\mathbf{\xi
,\qquad }\dot{x}^{k}\equiv \frac{dx^{k}}{d\tau }  \label{a5.12}
\end{equation}%
where $x^{k}=x^{k}\left( \tau ,\mathbf{\xi }\right) $. To obtain the quantum
description, we are to consider the statistical ensemble $\mathcal{E}\left[ 
\mathcal{S}_{\mathrm{st}}\right] $ of free stochastic relativistic particles 
$\mathcal{S}_{\mathrm{st}}$, which is the dynamic system described by the
action%
\begin{equation}
\mathcal{A}_{\mathcal{E}\left[ \mathcal{S}_{\mathrm{st}}\right] }\left[ x,u%
\right] =-\int m_{\mathrm{eff}}c\sqrt{g_{ik}\dot{x}^{i}\dot{x}^{k}}d\tau d%
\mathbf{\xi ,\qquad }\dot{x}^{k}\equiv \frac{dx^{k}}{d\tau }  \label{a5.14}
\end{equation}%
where $x^{k}=x^{k}\left( \tau ,\mathbf{\xi }\right) $, $u^{k}=u^{k}\left(
x\right) $, $k=0,1,2,3$. Here the effective mass $m_{\mathrm{eff}}$ is
obtained from the mass $m$ of the deterministic (classical) particle by
means of the change%
\begin{equation}
m^{2}\rightarrow m_{\mathrm{eff}}^{2}=m^{2}\left( 1+g_{ik}\frac{u^{i}u^{k}}{%
c^{2}}+\frac{\hbar }{mc^{2}}\partial _{k}u^{k}\right)  \label{a5.15}
\end{equation}%
where $u^{k}=u^{k}\left( x\right) $ the mean value of the 4-velocity
stochastic component. Using the change of variables%
\begin{equation}
\kappa ^{k}=\frac{m}{\hbar }u^{k},  \label{a5.16}
\end{equation}%
it is convenient to introduce the 4-velocity $\kappa =\left\{ \kappa ^{0},%
\mathbf{\kappa }\right\} $ with $\mathbf{\kappa }$, having dimensionality of
the length. The action takes the form%
\begin{equation}
\mathcal{A}_{\mathcal{E}\left[ \mathcal{S}_{\mathrm{st}}\right] }\left[
x,\kappa \right] =-\int mcK\sqrt{g_{ik}\dot{x}^{i}\dot{x}^{k}}d\tau d\mathbf{%
\xi ,}  \label{a5.17}
\end{equation}%
\begin{equation}
K\mathbf{=}\sqrt{1+\lambda ^{2}\left( g_{ik}\kappa ^{i}\kappa ^{k}+\partial
_{k}\kappa ^{k}\right) }  \label{a5.18}
\end{equation}%
where $\lambda =\frac{\hbar }{mc}$ is the Compton wave length of the
particle and the metric tensor $g_{ik}=$diag$\left\{ c^{2},-1,-1,-1\right\} $%
. In the nonrelativistic approximation this action turns into the action 
\begin{equation}
\mathcal{A}_{\mathcal{S}_{\mathrm{st}}}\left[ \mathbf{x,u}\right] =\int
\left\{ -mc^{2}+\frac{m}{2}\left( \frac{d\mathbf{x}}{dt}\right) ^{2}+\frac{m%
}{2}\mathbf{u}^{2}-\frac{\hbar }{2}\mathbf{\nabla u}\right\} dtd\mathbf{\xi }
\label{a5.19}
\end{equation}%
which coincides with the action (\ref{a5.4}) and generates the Schr\"{o}%
dinger equation for the irrotational flow of the fluid, described by this
action. In the relativistic case we obtain the Klein-Gordon equation.

Thus, we see that the proper modification of the space-time geometry
generates the multivariant dynamics, which describes quantum phenomena 
\textit{without\ any}\ \textit{additional\ suppositions.} It means, that we
obtain a \textit{fundamental theory}$\mathit{,}$ which replaces the
curtailed theory (conventional nonrelativistic quantum mechanics). Having a
fundamental theory, we may hope to construct the relativistic quantum theory
without any additional hypotheses.

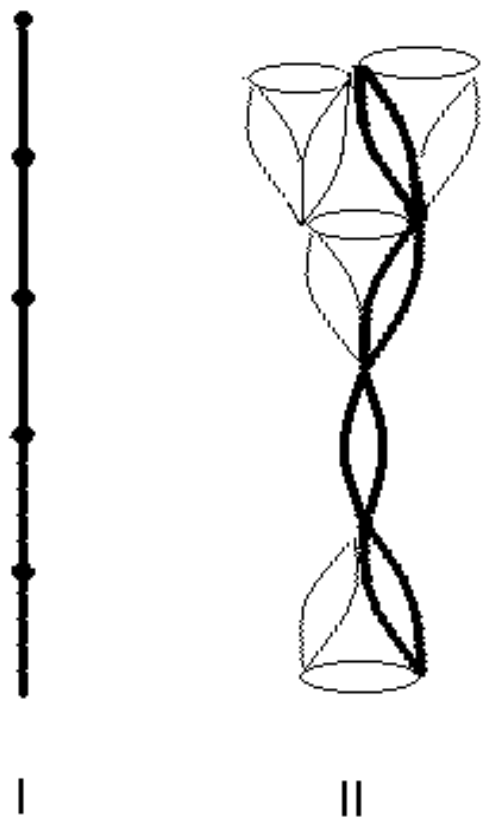
\end{document}